\documentclass[12pt,twoside]{article}
 \usepackage{fleqn,espcrc1}
\usepackage{graphicx}
 \begin{document} 

   \title{{\hfill {\small IFUP-TH 34/2000}}\\ \large Comparison of Transfer-to-Continuum and Eikonal Models of
   Projectile Fragmentation Reactions}
   \author{ A. Bonaccorso$^{(a),(b)}$ and G.F. Bertsch$^{(a)}$\\
    $^{(a)}$Dept. of Physics and Inst. Nuclear Theory\\
   University of Washington, Box 351550\\
    Seattle, WA 98195\\
    $^{(b)}$Istituto Nazionale di Fisica Nucleare, Sez. di Pisa, \\
    56100 Pisa, Italy}
   \maketitle

   \begin{abstract}

   Spectroscopic properties of nuclei are accessible with projectile
   fragmentation reactions, but approximations made in the reaction
   theory can limit the accuracy of the determinations. We examine
   here two models that have rather different approximations for 
   the nucleon wave function, the target interaction, and 
   the treatment of the finite duration of the reaction.  The
   nucleon-target interaction is treated differently in the eikonal and
   the transfer-to-continuum model, but the differences are more significant
   for light targets. We propose a
   new parameterization with that in mind. We also propose a new
   formula to calculate the amplitude that combines the better treatment
   of the wave function in the eikonal model with the better treatment of
   the target interaction in the transfer-to-continuum model.
    
   \end{abstract}

\begin{flushleft} 
 {\bf PACS }
number(s):25.70.Hi, 21.10Gv,25.60Ge,25.70Mn,27.20+n
 \end{flushleft}

\begin{flushleft}  {\bf Key-words}
Breakup, absolute cross sections, eikonal approximation, optical potential,
S-matrix, T-matrix.
 \end{flushleft} 
\newpage
   \section{Introduction}
   Heavy ion reactions at intermediate energy offer great promise
   to measure spectroscopic properties of nuclei far from 
   stability, but one needs a tractable reaction theory to interpret
   the experiments. In this respective the availability of higher
   energy heavy ion beams is most welcome, because it becomes
   a reasonable theoretical approximation to neglect exchange
   of nucleons between the colliding nuclei. One can therefore
   consider the interaction in each nucleus as that of an
   external (complex) potential field. Within the framework
   of this basic approximation and with given potentials, we will 
   here address the question of the accuracy of further simplified
   models of the reaction cross sections.
   A number of theoretical models have been proposed and
   calculated \cite{yab}-\cite{abfc} in
which
    different approximations were made. 
   In this work we focus on two of the models, the eikonal model
\cite{yab}-\cite{flo} and 
   the transfer-to-the-continuum (TC) model \cite{bb}-\cite{abfc}.

   There are several cross sections that are measured and calculated in
   the models. The simplest measurement is the single-neutron removal
   cross section, in which only the projectile residue, namely the
   core with one less nucleon, is observed in the final state. Besides
   the integrated removal cross section, denoted by $\sigma_{-n}$, the
   differential momentum distribution $d^3\sigma_{-n}/d k^3$ is also
   measured. A particularly useful cross section is $d \sigma_{-n} /d
   k_z$, the removal cross section differential in longitudinal momentum.
   If the final state neutron can also be measured, the corresponding
   coincident cross section $A_p \rightarrow (A_p-1) +n$ is called the
   diffractive breakup cross section. The difference between the removal
   and diffractive breakup is called the stripping cross section.

   This paper is organized as follows.
   In Sec. II we summarize the essential ingredients of the TC model and 
   the eikonal model, and discuss the accuracy of the neglect of finite
   interaction times in the eikonal model. In Sect III we discuss the
   different treatment of the neutron-target S-matrix in the two models.
   We will not discuss the accuracy of the Hankel function approximation
   separately, but it of course plays a role in the comparison of
   cross sections that we make in section IV which contains also our conclusions.

   \section {Theoretical models}
   All theoretical methods used so far rely on a basic approximation to describe the collision with
   only the three-body variables of nucleon coordinate, projectile
   coordinate, and target coordinate. Thus the dynamics is controlled by the
   three potentials describing nucleon-core, nucleon-target, and core-target
   interactions. In most cases the projectile-target relative motion is treated
   semiclassically by using a trajectory of the center of the projectile
   relative to the center of the target ${\bf s}(t)={\bf b_c}+{\bf v}t$
   with constant velocity $v$ in the $z$ direction and impact parameter
   {\bf b$_c$} in the $xy$ plane. Along this trajectory the amplitude for a
   transition from a nucleon state $\psi_i$ to a state $\psi_f$ is given by 
   \begin{equation}A_{fi}={1\over i\hbar}
   \int_{-\infty}^{\infty}dt<\psi_{f} (t)|V_{nt}({\bf
   r})|\psi_{i}(t)>,\label{1}\end {equation} where $V_{nt}$ is the
   neutron-target interaction. The state $\psi_i$ will be the bound state
   of the nucleon in the projectile, while the final state $\psi_f$ is a
   continuum state. The detailed derivation of Eq.(\ref{1}) from a scattering amplitude
containing the full time dependent propagator can be found in Sec.II of Ref. \cite{bb1}.
There it was shown to hold under the hypothesis that the breakup process is
limited to peripheral projectile-target trajectories and that it is   due mainly to the
neutron interaction with the target potential.   
  The probabilities for different processes can be
   represented in terms of the amplitude as \begin{equation} {dP\over d
   \xi}=\sum |A_{fi}|^2 \delta(\xi-\xi_f),\label{2}\end{equation} where
   $\xi$ can be momentum, energy or any other variable for which one measures
   a differential cross section.

   The effects associated with the core-target interaction will be included by 
   multiplying the above probability by the $b_c$-dependent probability for the core to be
   left in its ground state.

   Thus the differential cross sections with respect to the longitudinal momentum is 
    \begin{equation}
    {d\sigma\over d {k_z}}=C^2S
    \int_0^{\infty} d{\bf b_c} {d P_{-n}(k_z,b_c)\over dk_z} 
    P_{ct}(b_c), \label{cross} \end{equation}
   where $k_z$ is the (longitudinal) recoil momentum of the neutron
    (see Eq. (2.3) of \cite{me}) and $C^2S$ is the spectroscopic
   factor for the initial single particle orbital. The cross section can
   be further divided into a stripping cross section $\sigma_{str}$ and a
   diffractive breakup cross section $\sigma_{diff}$ depending on whether
   the removed neutron is detected in the final state or not. We
   accordingly will consider these individual probability distributions,
   \begin{equation} P_{-n} = P_{str}+ P_{diff},
   \label{4}\end{equation}
   and use a similar notation for the cross sections. 

   We first summarize the transfer-to-the-continuum (TC) model. This model
   treats the time-dependence of the reaction explicitly, thus conserving
   energy. It uses the on-shell neutron-target scattering
   matrix, therefore making it in principle model-independent with respect
   to that interaction. It also makes use of the asymptotic form of
   the neutron wave function in the projectile. This is an asset in
   that the formulas have an analytic limit, but a disadvantage in that the
   results are only reliable at peripheral impact parameters\footnote{
   The derivation of the TC model requires that the neutron-target and
   the neutron-projectile potentials do not overlap. In this respect
   the result that can be expressed entirely in terms of asymptotic properties reminds
   one of B\'eg's theorem\cite{be61}.}. The two breakup probabilities are
   given by the following expressions:
   \begin{equation}
   {dP_{diff}(b_c)\over d k_z} = \sum_{l_n} |1-S_{l_n}|^2 B(l_n,k_z,b_c),
   \label{5}\end{equation}
   \begin{equation}
   {dP_{str}(b_c)\over d k_z} = \sum_{l_n} (1-|S_{l_n}|^2) B(l_n,k_z,b_c),
   \label{6}\end{equation}
   where the factor $B(l_n,k_z,b_c)$ is a transfer
   probability which depends on the details of the initial and final
   states, and on the energy of relative motion. 
   It is given by
   \begin{equation} B(l_n,k_z,b_c)={1\over 2}\left({\hbar\over mv}
   \right){1\over k_f}(2l_n+1) \vert C_i\vert
   ^2{e^{-2\eta b_c} \over 2\eta b_c} M_{l_nl_i},\label{7}
    \end{equation}
    where $l_i$ is the angular momentum of the bound neutron in the
   initial state with respect to the core. The variable $l_n$
   has the interpretation as the angular momentum of the neutron
   with respect to the target. Also 
   \begin{equation}M_{l_nl_i}={1\over\sqrt{\pi}}\int_0^\infty dx e^{-x^2}
   P_{l_i}(X_i+B_ix^2)P_{l_n}(X_f+B_fx^2),\label{8}\end{equation}

   The arguments of the Legendre polynomials $P_{l_i}$ and
   $P_{l_n}$ are 
   $X_i=1+2(k_z/\gamma_i)^2$ and $X_f=2((k_z-m v)^2/(\gamma_i^2+2m v k_z
   -(mv)^2)-1$ , $B_i={2\eta/ d\gamma_i}$ and
   $B_f={2\eta/ dk_f}$. 
    $\gamma_i$ is related to the initial state binding energy
   by $\gamma_i=\sqrt{-2m \varepsilon_i}/\hbar$.
   The variable $\eta=\sqrt{k_z^2+\gamma_i^2}$ has the interpretation as
   the modulus of the transverse component of the
   neutron momentum. 
   In our notation
   $\varepsilon_f$ is the energy of the neutron relative to the target in
   the final state. For diffraction this is the same as
    the final laboratory energy of the neutron if the target
   recoil kinetic energy is neglected. In the case of stripping, 
   if it goes through compound
   nucleus formation, $\varepsilon_f$ is the excitation energy
   of the compound state above the neutron threshold in the
   residual nucleus. For inelastic scattering 
    it is the energy of the breakup neutron before it
   scatters from the target. This is equivalent to the sum of
   the excitation energy of the target final state and the
   final neutron energy relative to the target. If the target
   recoil kinetic energy is neglected the final kinetic energy
   $E_f$ of the ejectile is given by the energy conservation
   condition

    \begin{equation} E_f-E_{inc}= Q 
=\varepsilon_i-\varepsilon_f,\label{ec}\end{equation}
   where $E_{inc}$ is the initial incident energy of the
   projectile in the laboratory, Q is the reaction $Q$ value
   given by
   $Q=\varepsilon_i-\varepsilon_f$ and $\varepsilon_i$ is the
   initial neutron binding energy in the projectile. With
   this approximation Eq. (\ref{ec}) relates
   $k_z$ to the the projectile residue parallel momentum.
   Finally, $\vert C_i\vert^2$ is the asymptotic normalization constant of the
   initial bound wave function 
   \begin{equation}\psi_i(r)=-i^lC_i\gamma_ih_{l_i}^{(1)}(i\gamma_ir)
   Y_{l_im_i}(\theta,\phi), \,\,\,\gamma_i r >>1.
   \label{in}
   \end{equation}
   It is obtained by fitting a realistic radial wave function to the Hankel
   form $h_{l_i}^{(1)}$ outside the potential radius. In this way the
   transfer to the continuum results are model dependent. On the other hand
   the dimensionless quantity $\Lambda_{l_i}=\gamma_i^{-1}\vert C_i\vert^2C^2S$ 
   has been called reduced normalization in the contest of
   spectroscopy done with transfer reactions \cite{sat,ker}. If one
   considers $\Lambda_{l_i}$ as the prefactor of the theoretical cross section
   obtained with the Hankel function, the ratio between the
   experimental cross section and the theoretical cross section would
   determine its value.

   Eqs.(\ref{5}-\ref{7}) were derived assuming no overlap between
   neutron-core and neutron-target potential. This assumption can be
   avoided if one makes an eikonal approximation to the basic
   expression for the amplitude, Eq.(\ref{1}), as shown in Appendix B. Then one
derives
\cite{ab,abb}
   \begin{equation}{dP_{-n}(b_c)\over d k_z}\sim {1\over 2\pi}
   \int_0^\infty d{\bf b_n} \left[|(1-e^{-i\chi({ b_n})})|^2 +
    1-|e^{-i\chi({
   b_n})}|^2\right]
    |\tilde {\psi}_i ({\bf b_n-b_c}, k_z)|^2 ,\label{pg}\end{equation}
   where $\bf b_n$ is the transverse coordinate of the neutron with respect to
   the target. The neutron-target $S$-matrix is approximated by the
   eikonal form $\bar S(b_n)= e^{-i\chi({ b_n})}$, related to the
   optical potential $V_{nt}$ by 

\noindent $\chi({\bf b_n})={1\over \hbar v}\int_{-\infty}^{\infty} 
   V_{nt}(x,y,z^{\prime})dz^{\prime}$. Finally,
   \noindent $|\tilde {\psi}_i({\bf b_n-b_c}, k_z)|^2$ is the
   longitudinal Fourier transform of the initial state wave function. 
    Also the connection between
   Eqs.(\ref{5}-\ref{7}) and Eq.(\ref{pg}) is made by replacing the sum over
    partial waves in Eqs.(\ref{5},\ref{6}) by the integral over impact
   parameters as in Eq.(\ref{pg}), and by evaluating the longitudinal Fourier
   transform in Eq.(\ref{pg}) using the asymptotic initial state wave function,
   Eq.(\ref{in}). The one-dimensional Fourier transform of the initial wave
   function can be calculated analytically in the case of an Hankel function
   approximation Eq.(\ref{in}) yielding
   \begin{eqnarray}
   {1\over (2l_i+1)}\sum_{m_i}| \tilde {\psi}_{l_im_i}({\bf
   b_c-b_n},k_z)|^2
     &=&
   {1\over (2l_i+1)} \sum_{m_i}|2 C_iY_{l_i,m_i}
   (\hat k_z) K_{m_i}(\eta \rho)|^2\nonumber \\ &\approx &
   C_i^2 {e^{-2\eta \rho}\over 2\eta \rho}P_{l_i}(X_i)
   , \label{2a} \end{eqnarray} 
   where $\rho=\vert {\bf b_c-b_n}\vert $.

   The total breakup probability is obtained from the integral of
   Eq.(\ref{pg}) involving

   \begin{equation}
   I(k_z^{min},k_z^{max}) =\int_{k_z^{min}}^{k_z^{max}}d k_z \vert\bar \psi_i({\bf
   b_n-b_c},k_z)\vert^2.\label{phsp}
   \end{equation}
   $k_z^{min}$ and $k_z^{max}$ are the kinematically allowed minimum and maximum neutron
   parallel momenta discussed in the following. Therefore although Eq.(\ref{pg})  
describes the neutron-target rescattering in the eikonal approximation, it still satisfies
 neutron energy and momentum conservation. 

On the other hand, a sudden approximation or frozen halo form of the eikonal
method has been used in \cite{yab,esb1,jef} where energy conservation was 
neglected. This approximation can be derived from Eq.(\ref{pg}) when the integration
limits in Eq.(\ref{phsp}) can be extended to
$\pm\infty$. Then Eq.(\ref{phsp})  is just be the longitudinal
   density, and the formulae for the TC and eikonal model  become
   identical. In fact in this limit
   the removal cross section reduces to
   \begin{equation}\sigma_{-n}=C^2S
   \int d^2{\bf b_c}\int d^3{\bf r_n}\left[|(1-\bar S)|^2 + 1-|\bar
   S|^2\right] |S_{ct}(b_c)|^2|{\psi}_i ( {\bf
   b_n-b_c},z)|^2,\label{creik}\end{equation}
    which is consistent with the 
   breakup cross section of \cite{yab,esb1}. In Eq.(\ref{creik}) $\bar S$ is an eikonal
S-matrix, as defined after  Eq.(\ref{pg}).

   The steps necessary to obtain Eq.(\ref{creik}) from Eq.(\ref{pg}) and
   Eqs.(\ref{4}-\ref{6}) can be justified in the high energy limit as follows. If the
   neutron binding energy is not too large, the final energy or momentum distributions are
   strongly peaked at the incident energy per nucleon. Therefore it is possible to average
   the dependence of the neutron target optical model S-matrix of Eqs.(\ref{5},\ref{6})
   over the full range of neutron continuum energies and assume that the S-matrix can be 
   approximated with the eikonal values $\bar S$
     obtained at $\varepsilon_f={1\over 2} mv^2$. On the other hand, for the
   eikonal approximation to be good it is necessary that the parallel
    component of the neutron momentum $k_z$ be large with respect to the 
   transverse component $\eta$.
    In particular such condition must be 
   satisfied for
   the minimum values of both. 
   From the definition of $k_z$ it is easy to see that its lowest possible
   value is $k_z^{min}=-(\varepsilon_i+{1\over 2}mv^2)/(\hbar v)$
    in correspondence to $\varepsilon_f=0$. And that the minimum value of
   $\eta=\gamma_i$. Therefore we get $|k_z^{min}|>>\gamma_i$ if
   ${\sqrt E_{inc}}/2>>\sqrt{\varepsilon_i}$. 
   For a real halo with separation energy around 0.5MeV the condition
   $|k_z^{min}|>>\gamma_i$ is satisfied at all initial energies. Increasing the binding
   energy it is necessary to go to higher incident energies in order for the parallel
   momentum component to be larger than the transverse one. For a typical binding
   of 10MeV or more the conditions for the sudden eikonal approximation to be valid
   are satisfied from about 80A MeV.
   It is useful also to consider the values of 
    neutron-target center of
   mass momentum $k_v=\mu v/\hbar$ where $\mu=A_t/(A_t+1)$ is the neutron-target reduced
   mass. Here too we see that for a typical diffuseness of a=0.5fm the semiclassical
   condition $ak_v>>1$ is satisfied starting from about 80MeV.
    Finally if such condition is satisfied we can extend the
   lower limit in the $k_z$ integral to $-\infty$. 
   There is however also an upper limit to the final neutron or ejectile
   momentum value in the parallel direction, due to the energy and momentum
   conservation. It was discussed in ref.\cite{huf} where it was
   shown that 
   \begin{equation} k_z^{max}=\left (1-{1\over
   2A_p}\right)m\label{kmax},\end{equation} m is the neutron mass.
   $k_z^{max}=0.96fm^{-1}$for a $^{12}Be$ projectile. In terms of the
   maximum neutron final energy in the continuum the above condition
   reads

   \begin{equation} \varepsilon_f^{max}=k_z^{max}\hbar
   v+{1\over 2}mv^2+\varepsilon_i.\end{equation}

   If this upper limit is also extended to $+\infty$ then
    the $k_z$ integral gives a $\delta$-function and inserting
   Eq.(\ref{pg}) and the damping factor $P_{ct}=|S_{ct}|^2 $
    in Eq.(\ref{cross}) we finally get Eq.(\ref{creik}).

   These very strict conditions necessary to extend the theoretical limits of the $k_z$
   integral to infinity, can be somewhat revised in practical calculations. To see
   how accurate the sudden approximation is, we have calculated the
   integral $I(k_z^{min},k_z^{max})$ under various conditions of angular momentum, neutron
   binding energy in the projectile, and projectile velocity. For values of the parameters
   of interest there can be a rather large reduction for small values of the
   neutron transverse radius in the projectile, $|b_n-b_c|$. However,
   the approximation becomes increasingly accurate as the transverse
   radius is made larger.
   In Fig. 1 we show the regions of energies in which the ratio
   $$
   R={I(-\infty,\infty)\over I(k_z^{min},k_z^{max})},
   $$
   is within 5\%, 5-10\% and 10\% of being unity, for a $d$-wave orbital at a transverse radius corresponding
   to strong absorption. The region of good
   agreement is somewhat larger for $s$- and $p$-waves. Thus it 
   seems that the sudden approximation should be acceptable for weakly
   bound neutron orbitals ($< 6MeV$) and beam energies greater than 40A MeV, provided the
   reaction samples the initial wave function at distances rather larger than the sum of
   the radii of the reaction partners.

    It must be noted that here
   as in Eq.(\ref{cross}) we have adopted the so called no-recoil
   approximation \cite{bhe} which consists in 
    factorizing out from the matrix element the core-target S-matrix.
   Another important remark has to do with the extension of the 
   lower limit in the $k_z$ integral to $-\infty$. Values of $k_z$ smaller
   than $k_z^{min}$ obtained when $\varepsilon_f=0$ would correspond to
   final states in the projectile below the breakup threshold for the
   diffraction term. For the stripping term they would correspond to final
   bound states in the target. Taking $k_z^{min}\rightarrow -\infty$ is
   analogous to the completeness relation introduced in
   \cite{yab}. For this reason in Eq.(\ref{creik}) the diffraction term has
   usually been corrected by subtracting terms like $|<{\psi}_i|1-\bar S
   S_{ct}|{\psi}_i>|^2$. In the transfer to the continuum model this
   correction is not as important because the neutron final energy is
   always positive and the $k_z$ integration is performed in the
   kinematically allowed region.

    Estimates of absolute breakup cross sections
   published so far have been made either using Eq.(\ref{cross}) with
   the breakup probability given by the transfer to the continuum
   theory Eqs.(\ref{4}-\ref{8}) \cite{abb,me,abfc}
    or with Eq.(\ref{creik}) \cite{yab,esb,jef,flo,be12}.
   There are a number of assumptions contained in both models,
    the validity of which we are going to study in this
   paper.

     \section{n-target optical potential and S-matrix}

   Experimentalists often use light targets such as $^{9}Be$ or
   $^{12}C$ to study spectroscopy by projectile fragmentation
   reactions. The definition of an optical potential for light targets
   is a very delicate issue which has been discussed in the literature
   for long time \cite{jlm}. There are two main issues, i)
   global parameterizations are based on large nuclei; ii) the
   imaginary optical potential changes drastically for light nuclei.
    Recently these problems have been faced in the contest of
   halo breakup studies \cite{jef,me,abfc} where different choices have
   been made by different authors. Bertsch et al.\cite{esb} used the
   Varner et al. parameterization \cite{var}. Such parameterization has
   been obtained taking into account relatively low energy proton and 
   neutron cross sections (10-26MeV) on heavy targets, A=40-209. In
   ref.\cite{jef} a neutron target S-matrix was constructed in the
   optical limit of Glauber theory. The corresponding optical
   potential does not need to be given explicitly, but the behavior
   of the transmission coefficient given in \cite{jef} would suggest a
   strongly absorbing potential of volume type. Bonaccorso \cite{me}
   extended to high energies a parameterization of a phenomenological
   n-$^{9}Be$ optical potential fitted to low energy data. This
   potential seemed however to overestimate the free particle cross
   sections at high energies \cite{me} and therefore ref.\cite{abfc}
   also considered a microscopic optical
   potential calculated according to the method of Jeukenne, Lejeune and Mahaux
   (JLM) 
   \cite{jlm}. The JLM potentials are more complicated to calculate
   and to use than a simple standard parameterization. Then for the
   purpose of the present paper we have attempted to fit the available
   experimental n-$^{9}Be$ total cross sections with a new potential
   of the standard form. The potential is given by a real Woods-Saxon
   well and it has both volume and a surface derivative Woods-Saxon
   forms for the imaginary part. In obtaining the parameters given in
   Table I we have been guided by some existing parameterizations
   \cite{pot1,wat} obtained by fitting low energy data. We have
   modified such parameterizations to get a smooth behavior
   of the free particle neutron-$^{9}Be$ cross section in the energy
   range 10-180 MeV. The free-particle cross section have been
   obtained in an optical model calculation. A
   good agreement of calculated free
    neutron angular distributions with the data of ref.\cite{pot1} has also been obtained.
   This potential satisfies very well the subtracted dispersion relations
   as given for example in Ref.\cite {ma1} at energies larger than 40MeV. For lower
   energies the agreement is less good at small radii.

   In Fig. 2, top left,
   we show the experimental cross sections \cite{fin} together with the optical model
   cross sections, where the full curve is the total cross section, the
   dot-dashed curve is the elastic cross section while the dashed curve is
   the reaction cross section. At the bottom left we show the same quantities
   calculated in the eikonal approximation. The effect of the eikonal
   approximation is to reduce the calculated cross sections, in particular
   at the lower energies. The reduction is more pronounced for the elastic
   cross section, probably because quantum mechanical reflection effects on
   the potential barrier can not be reproduced by the eikonal
   approximation. For comparison we show at the top right and bottom right respectively
    the
   same quantities calculated with the Varner parameterization. Varner potential
   does not reproduce the high energy data, also it gives a dominance of
   the elastic scattering up to about 80MeV. Here too the eikonal
   calculations underestimate the elastic scattering at low energies.

   Our potential has a rather strong surface term. It gives also large
   volume integrals ($J_W/A\approx 200MeVfm^{3}$) in accordance with the
   light nuclei systematics\cite{jlm,ph}. The microscopic origin of it can
   be understood since 
   $^{9}Be$ is weakly bound and it has rather high breakup 
     probability. Also 
    it has been known for long time \cite{love} to have an unusually
   large mean square radius and to be strongly deformed. Finally we notice that we get a very
   large elastic vs reaction cross section ratio at low energy. The JLM
   potential gives the same behavior. The little experimental free
   particle data available
    show similar trend and it would be interesting to see if
   breakup reactions around 20A MeV which will soon be feasible at GANIL,
   will parallel such a behavior.

   In Fig. 3 we show the behavior of the term $|1-S|^2$ and of the
   transmission coefficient $1-|S|^2$ when calculated by the optical model
   (solid and dashed line respectively) and by the eikonal approximation
   (dotted and dot-dashed line). The results are shown as a function of the
   angular momentum $l_n$ and in the case of the eikonal calculations the
   semiclassical relation $l_n+1/2=kb_n$ was used to make the connection between the
   angular momentum and the neutron impact parameter. We give results at energies ranging
   from 20 to 100 MeV. An interesting characteristic of these results is that the eikonal 
   seems to concentrate the scattering at larger impact parameters than in the optical model
   case. Such effects have been seen by other authors \cite{av} in different
   situations that ours and sometimes it has also been shown \cite{flo,av}
   that higher order eikonal corrections can improve the agreement with the
   quantum mechanical calculations. It is possible that in our case the
   effect is amplified by the deep surface part of the imaginary
   potential. It is well known that the eikonal approximation works well
   in the presence of strongly absorbing volume potentials and
   differences would also be less extreme for potentials that have
   more diffuse edges. This is because the reflection effects at the
   barrier that modify the neutron trajectory are not taken into
   account in a semiclassical approach. 

   From the differences seen here, we would expect
   quite large differences between predictions of the
   two treatments of the S-matrix. The quantification
   of these differences is the subject of 
   section IV.

   \section{Cross sections and conclusions}

   We start this section by showing some results for the total probability

   \noindent $P(b_c)=P_{-n}(b_c)P_{ct}(b_c)$ obtained from the integrand of
   Eq.(\ref{cross}) after integrating over $k_z$ and using
   Eq.(\ref{prs}). The
   $R_s$ values are given in Table III at each energy. They were obtained, as
explained in Appendix A,
   from the S-matrix calculated by folding the neutron-target
   optical potential of Table I with the projectile density, at all
   energies but 20A MeV. At such a low energy the optical limit of the
   eikonal model cannot be justified, but calculations by Carstoiu \cite{floc} in second
order eikonal approximation, folding the JLM
   potentials agree with our $R_s$ at all energies, including 20A MeV.
Same values are also obtained from the optical potential of \cite{jat}.

   For our numerical comparison of the various methods, we have chosen
   to study the breakup from
   the s and p states in $^{12}Be$ which were recently measured by
   Navin et al.\cite{be12}. The initial bound state parameters are given in
   Table II.

    In
   Fig. 4 we show the total probabilities for 
   diffraction obtained in the TC calculation by the solid
   lines. The dashed lines are for the absorption or stripping. Energies are
   as in Fig. 3. The dotted lines are the calculations from the
   diffraction integrand of Eq.(\ref{pg}) while the dot-dashed lines are for the
   stripping. Because of the shift in the S-matrices of Fig. 3, we see
   here that the eikonal calculations have their maxima at slightly larger
   core-target distances than the optical model calculations. At the lowest
   energy the optical model gives almost equal diffraction and stripping
   probabilities. The eikonal calculation gives a rather larger
   diffraction probability. In all other cases the stripping is the
   dominant term. This effect was first pointed out in normally bound
   nuclei \cite{tiina,bb4} and it has been confirmed by some experiments
   \cite{henri} and other theoretical calculations \cite{abfc}.

   The TC calculations discussed in this paper are
   performed using the asymptotic, Hankel form of the initial state wave
   function, which can be the origin of some divergence if the projectile
   and target potential have an important overlap during the reaction. This
   problem was discussed in great detail in Ref.\cite{bb1} where it was shown to be more
important for
   the stripping term since it is proportional to the neutron target phase shift, while the
   diffraction term, being proportional to the square of the phase shift converges more
   rapidly. This situation reflects the fact that diffraction reactions involve a  more
peripheral part of the initial wave function
   than stripping reactions.
    Here we can easily see from Eq.(\ref{pg}) that the integral over
   $b_n$, the
    neutron target impact parameter is well behaved provided the
   neutron-target potential and the corresponding S-matrix fall off more
   rapidly than the tail of the initial wave function. A practical way to satisfy such a
condition and to avoid any divergence in the sum over neutron partial waves in
Eq.(\ref{6}) is to use  a small value for the diffuseness of the imaginary part of the
optical model (cf. Sec.IV of \cite{bb1}). Otherwise, as it is shown in Appendix B it is
possible to start again from Eq.(\ref{1}) and obtain a formula which contains the
realistic single particle wave function and an optical model S-matrix, still keeping the
information on the proper kinematical limits for the neutron parallel momentum. The only
hypothesis involved is that high neutron angular momenta are dominant.

To show the amount of error that is introduced by the Hankel approximation in the TC
method, we show in
   Fig. (5) the integrand functions of the stripping and diffraction terms of
Eq.(\ref{pg}), after integrating over $k_z$, obtained
   from the realistic bound state wave function and the corresponding terms
   in the sum over partial waves of Eqs.(\ref{5},\ref{6}) in the
   case of the incident energy of 78A MeV. The calculations were done at
   the fixed impact parameter
   $b_c=5.6fm$ between the projectile and target. Left figure is for the diffraction term
   while right figure is for the stripping term.
    We can see that the calculation for stripping according to Eq.
(\ref{6}) which is done with the Hankel function and optical model S-matrix (diamonds)
 shows a
   slower decrease in the interior of the projectile ( small b$_n$ ) than the calculation
   with the realistic wave function (crosses) done according to Eq.(\ref{a10}) of Appendix
B. Therefore the comparison between these two results gives an indication of the effect of
the Hankel function approximation. On the other hand the eikonal calculations (full
curve), done with the full wave function
    have higher maxima which occur at a larger distance than the
   optical model calculation. This effect is
   due to the behavior at large impact parameters of the eikonal
   S-matrix, as we have discussed before. The comparison of the full curve with the
crosses gives an indication of the effect of the eikonal approximation. In the end the 
   integrated probabilities obtained with the three possible approximations contained in
Eqs. (\ref{5}) and (\ref{6}), Eq.(\ref{a10}) and Eq.(\ref{a7}) differ by not more than
about 10\%. This leads to about 20\%
   difference in the total cross sections because of the further integration over
   $b_c$, the core-target impact parameter. From the peak values of the diffraction and
   stripping curves it is easy to see that the stripping is almost three times larger than
   the diffraction.

   The values of the total cross sections are given in Table III.
    Because the eikonal
   probabilities are shifted towards larger distances we get total cross
   sections which are larger by about 20\% than the TC
   calculations at all energies but 20A MeV. At this lowest energy 
    two effects tend to compensate each other:
    the eikonal free particle cross section is largely
   underestimated while there is an overestimate of the
   phase space kinematically allowed. Then the eikonal and optical model
    calculations seem to give very close results.

    Considering all the effects discussed above
    we conclude than present
   models used to calculate the breakup cross section might
   tend to overestimate the true value, but the error should be of the
   order of 20\% or less.
    On the other hand presently available experimental data are inclusive 
   with respect to the target which was unobserved. In the case of
   $^{9}Be$ it is possible that some part of the cross section comes from
   reactions in which the target itself underwent breakup. These reactions
   could account for about 20\% of the measured cross sections and they are
   certainly not taken into account by presently available theoretical
   models. We suggest therefore that future experiments detect the final
   excitation state of the target in coincidence with breakup events (as
   done for example in Ref.\cite{henri} for "normal" heavy ions) and at the
   same time that theoretical models be improved through a more realistic
   treatment of the core-target interaction, as mentioned in Sec.IV and in Appendix A. We propose
also  an improved calculation of the breakup amplitude using the new formula
Eq.(\ref{a10}) derived in Appendix B
 that combines the better treatment
   of the wave function in the eikonal model with the better treatment of
   the target interaction in the transfer-to-continuum model. 
 We leave to a future study to apply this equation
   to the reaction data.

   Ejectile momentum distributions are not shown in this paper. They have been
   discussed in great detail in \cite{abb,me,abfc} where it was shown that
   there can be noticeable differences in the results of the TC and eikonal
   methods. Present data do not have enough statistics to distinguish clearly 
   between the two models. 
    
   {\bf Acknowledgments}

   This research was supported by the U.S. Department of Energy under Grant
   DE-FG03-00-ER41132. We wish to thank David Brink for several discussions and suggestions,
   Florin Carstoiu for communicating his results before publication and Juan Carlos Pacheco
   for checking that our n-target optical potential satisfies the dispersion relations.

   \appendix \section{ Ion-ion S-matrix}

   We discuss here the theory of the core-target S-matrix, needed to calculate
   $P_{ct}(b_c)$ in Eq.(\ref{cross}). The best source of information comes
   from the ion-ion reaction cross section which is related to the
   core-target S-matrix by \begin{equation}\sigma_R^{ct}=\int_0^{\infty
   }d^2{\bf b_c}\left (1-\vert S_{ct}({\bf b_c}) \vert
   ^2\right).\end{equation}

   Here we have assumed that the semiclassical replacement of the partial wave sum by an
   integral over ${\bf b_c}$ is permitted. Unlike the nucleon-nucleus
   scattering, this is always a good approximation, because of the smaller
   wave length involved, and one can safely calculate this S-matrix in the
   eikonal approximation. The usual procedure is to define an optical
   potential for the core-target scattering, and calculate the S-matrix
   from it. In practice the behavior of the S-matrix is determined by two
   parameters. The first and most important is the strong absorption radius
   $R_s$, defined as the distance of closest approach for a trajectory that
   is 50\% absorbed from the elastic channel. The reaction cross section
   practically speaking is determined by the strong absorption radius. The
   next most important parameter is the thickness $a$ of the absorption
   region. In optical model fits, it is closely related to the asymptotic
   behavior of the imaginary potential $W(r)\sim e^{(-r/a)}$. Just as the
   reaction cross section strongly constrains the strong absorption radius,
   measurements of the elastic angular distribution provide information
   about $a$.

   In this work we shall use a simple parameterized form \cite{me} for 
   $P_{ct}({\bf b_c})$, namely
   \begin{equation} P_{ct}(b_c) = \exp (-\ln 2 e^{ (R_s-b_c)/a } ) . \label{prs} 
   \end{equation} 

   It would be correct to use the elastic scattering S-matrix if the experimental cross
   sections were reported for projectile breakup leaving the target in its ground state.
   In fact the target final state is difficult to measure and most experimental cross
   sections are inclusive with respect to the target final state. Thus the measured cross
   sections should be somewhat larger than one would calculate with an elastic core-target
   S-matrix.

   A way to include absorption of the projectile core without regard to the target is to
   construct an optical potential convolution of a nucleon-ion potential with the target
   density. In the calculations of this work we have determined strong absorption 
   radii by this method, using the optical potential of Sec. III. We also take a
   diffuseness parameter $a=0.6fm$. 

 \section{ Breakup Amplitude}

   In this section we give a more general formulation of the breakup problem which is less
   dependent on the reaction model assumptions of the text. We start with
   Eq. (1), again assuming a straight-line trajectory. Using a Galilean
   transformation for the initial wave function as described in ref.\cite{bb1}, 
   Eq. (1) becomes

   \begin{equation}A_{fi}={1\over i\hbar v}
   \int d{\bf b_n} dz\phi_{f} ({\bf r})V_{nt}({\bf
   r})e^{ik_{f_z}z}\tilde \phi_{i}({\bf b_n-b_c}, k_z),\label{a2}\end {equation} 
   where
   $k_z=k_1=-(\varepsilon_i-\varepsilon_n+{1\over 2}mv^2)/(\hbar v)$ and
   $k_{f_z}=k_2=-(\varepsilon_i-\varepsilon_n-{1\over 2}mv^2)/(\hbar v)$ are the
    $z$-components of the neutron momentum in the initial and final state
   respectively.

   Introduce now a T-matrix in a mixed representation as

   \begin{equation}T({\bf k_n},{\bf b_n},k_2)= {1\over i\hbar v}\int dz\phi_{f} 
({\bf b},z)V_{nt}({\bf
   b},z)e^{ik_{2}z},
   \label{a3}\end{equation} where ${\bf k}_n$ is the neutron final (measured) momentum.

   If the eikonal approximation for the final wave function is used

   \begin{equation} \phi_{f}^* ({\bf b},z)=e^{-i{\bf
   q}_\perp\cdot {\bf b_n}}e^{-iq_zz}e^{{i\over \hbar
   v}\int_{+\infty}^z V_{nt}(x,y,z^{\prime})dz^{\prime}} ,\label{a4}\end{equation} 
    the $dz$ integral in Eq.(\ref{a3}) can be calculated by parts when
   the initial binding energy is small and the final neutron energy is 
   close to the neutron incident energy and the neutron scattering angle is small
   such that $k_2-q_z \simeq 0$. Then

   \begin{equation}{i\over \hbar v}\int dz e^{i(k_2-q_z)z}V_{nt}({\bf
b_n},z)e^{{i\over \hbar v}
   \int_{-\infty}^z e^{i(k_2-q_z)z^{\prime}}V_{nt}(x,y,z^{\prime})dz^{\prime}}=
   1-e^{-i\chi({\bf b_n})},\label{a5}\end{equation}
   where
   \begin{equation}\chi({\bf b_n})={1\over \hbar v}\int_{-\infty}^{\infty} 
   e^{i(k_2-q_z)z^{\prime}}V_{nt}(x,y,z^{\prime})dz^{\prime}\approx 
   {1\over \hbar v}\tilde V_{nt}({\bf b_n},0),\label{a6}\end{equation}
   where $V_{nt}$ is a complex potential whose real and imaginary
   strengths are negative. We finally obtain the breakup amplitude
   in the eikonal 
   form as \begin{equation}
   A_{fi}=\int_0^{\infty}d{\bf b_n}e^{-i{\bf q}_\perp\cdot{\bf b_n}}
   \left(1-e^{-i\chi({\bf b_n})}\right )\tilde \phi_{i}({\bf b_n-b_c}, k_z)
   .\label{a7}\end{equation}

   Similarly a more general T-matrix in a mixed representation can be obtained starting from the partial
   wave form 
   \begin{equation}T({\bf k_n},{\bf k}_{\perp},k_2)
   ={\hbar^2\over2m}{4\pi\over 2ik_n}\Sigma_{l_n}(2l_n+1)(S_{l_n}-1)P_{l_n}(\hat{\bf
   k}_n\cdot \hat{\bf k}_{\alpha}),\label{a8}\end{equation}
   on the energy shell such that $k_n=k_{\alpha}$, and defining

   \begin{eqnarray}T({\bf k_n},{\bf b_n},k_2)&=&\int {d{\bf k}_{\perp}\over (2 \pi)^2} e^{-i{\bf
   k}_\perp\cdot{\bf b_n}}T({\bf k_n},{\bf k}_{\perp},k_2) \nonumber \\ &=&
   {\hbar^2\over2m}{4\pi\over 2ik_n}\Sigma_{l_n}(2l_n+1)(S_{l_n}-1)\nonumber \\&& \times \int {d{\bf
   k}_{\perp}\over (2
   \pi)^2} e^{-i{\bf k}_{\perp}\cdot{\bf b_n}}P_{l_n}(\hat{\bf k}_n\cdot \hat{\bf
   k}_{\alpha}).\label{a9}\end{eqnarray}
   This is unfortunately off shell but we will do the following high energy
   approximations similar to what we have done above in the eikonal case. In
   the large angular momentum, small scattering angle limit, define 
   $b_n^{\prime}={(l_n+{1\over 2})\over k_n}$,
     we can substitute the partial wave sum
   with an integral and also use $P_{l_n}(\hat{\bf k}_n\cdot \hat{\bf
   k}_{\alpha})\rightarrow J_0(|{\bf
   k}_n-{\bf k}_{\alpha}|b^{\prime})=(2\pi)^{-1}\int d\phi e^{-i({\bf q}_\perp-{\bf
   k}_\perp)\cdot {\bf b_n^{\prime}}}$ where ${\bf q}_{\perp}$ is the transverse
   component of ${\bf k}_n$ . Then

   \begin{eqnarray}A_{fi}&=&{1\over i\hbar v}\int d{\bf b_n}\int{d{\bf k}_{\perp}\over (2
   \pi)^2}e^{-i{\bf k_{\perp}\cdot b}}T({\bf k_n,k}_{\alpha}) {\tilde
   \phi}_{i}({\bf b_n-b_c},k_1)\nonumber \\
   &=&{1\over i\hbar v}{\hbar^2\over2m}{4\pi\over
   2ik_n}\Sigma_{l_n}(2l_n+1)(S_{l_n}-1)\nonumber \\&&
   \times
    \int d{\bf b_n}\int {d{\bf k}_{\perp}\over (2 \pi)^3}\int d\phi e^{-i({\bf
   q}_\perp-{\bf k}_\perp)\cdot {\bf b_n^{\prime}}}e^{-i{\bf k}_\perp\cdot{\bf
   b}} \tilde \phi_{i}({\bf b_n-b_c},
   k_z)
   \nonumber \\&\approx& -{\hbar k_n\over mv}\int d{\bf b_n^{\prime}}e^{-i{\bf
   q}_\perp\cdot {\bf b_n^{\prime}}}(S_{l_n}-1)\int d{\bf b_n}\int {d{\bf
   k}_{\perp}\over (2
   \pi)^2} e^{-i{\bf k}_\perp\cdot ({\bf b_n-b_n^{\prime}})}\tilde \phi_{i}
   ({\bf b_n-b_c},
   k_z)
   \nonumber \\&=&\int d{\bf b_n}e^{-i{\bf q}_\perp\cdot {\bf b_n}}
   (1-S_{l_n})\tilde \phi_{i}({\bf b_n-b_d}, k_z).\label{a10}\end{eqnarray}

   This formula combined the improved treatment of the interaction by the
   TC theory with the better treatment of the neutron wave function by
   the eikonal theory.

 \newpage

 \setlength{\oddsidemargin}{-1cm}
   \noindent{\bf Table I.}
   Energy dependent optical model parameters.
   a$_R$=0.387fm, r$_I$=1.368fm, a$_I$=0.3fm at all energies.
   \begin{center}
   \begin{tabular}[bht]{|ccccc|} \hline\hline
   $\varepsilon_f$&$V_R$&$r_R$&$W_S$&
   $W_V$\\
   (MeV)&(MeV)&(fm)&(MeV)&(MeV)\\&&&&\\
   20-40&$38.5-0.145\varepsilon_f$&$1.447-0.005(\varepsilon_f-20)$&
   $1.666+0.365\varepsilon_f$&$0.375\varepsilon_f-7.5$\\ 
   40-120&&&$16.226-0.1(\varepsilon_f-40)$&$7.5-0.02(\varepsilon_f-40)$\\ 
   120-180&&&$8.226-0.07(\varepsilon_f-120)$&5.9\\ 
          \hline
   \end{tabular}
   \vskip 1cm
   \end{center}
   \noindent{\bf Table II.}
   Initial state parameters in $^{12}$Be. Binding energies in MeV,
   asymptotic normalization constants $C_i$ in fm$^{-{1\over 2}}$.
   \begin {center}
   \begin{tabular}[bht]{|ccccc|} \hline\hline
   $J,
   \pi$ &$l$&$j$&{$|\varepsilon_i|$}&{$C_i$}\\
   \hline
   $1/2^+$& 0&0.5 &3.32&2.45\\
   $1/2^-$& 1&0.5 &3.52&1.29\\ 
   \hline
   \end{tabular}
\end{center}
   \vskip 1cm

   \noindent{\bf Table III.}
   Breakup cross sections in mb from the initial states 
   $1/2^+$ and $1/2^-$ in $^{12}Be$ on $^{9}Be$, at several incident
   energies. Incident energies inA MeV, $R_s$ in fm. Experimental data
   and shell model spectroscopic factors from \cite{be12}.
 \begin{center}
   \begin{tabular}[bht]{|ccccccccccccc|} \hline\hline
   $J,\pi $&$E_{inc} $ &$R_s$& $\sigma_{str}$ & $\sigma_{str }$& 
   $\sigma_{diff}$&$\sigma_{diff }$ & $\sigma_{-n}$& 
   $\sigma_{-n}$&$\sigma_{exp}$&$S_{SM}$&$S_{TC}$&$S_{eik}$\\ 
   &&&TC&eik&TC&eik&TC&eik&&&&\\
   \hline
   $1/2^+$&20& 6.5& 35&31&45&46&80&77&&&&\\
   &40&6.1&41&51.6&31&35&72&86.6&&&&\\
   &60&5.8&43.5&54.6&24&31&67.5&85.6&&&&\\
   &78&5.6&43&54.8&19&24.7&62&79.5&32(5)&0.69&0.65&0.51\\
   &100&5.4&42&53&13&17.6&55&70.6&&&&\\
   \hline
   $1/2^-$&20& 6.5& 19&16.5&21&22.7&40&39.2&&&&\\
   &40&6.1&23.4&29.6&16.3&18&39.7&47.6&&&&\\
   &60&5.8&27.2&32.7&13&15.9&40.2&48.6&&&&\\
   &78&5.6&28&34&11&13&39&47&18(3)&0.58&0.54&0.45\\
   &100&5.4&28.4&33.7&8.4&9.7&36.8&43.4&&&&\\
   \hline
   \end{tabular}
   \end{center}
   \vskip 1cm
  
   \newpage

   {\bf Figure Captions.}

   \vskip .2in
   \noindent {\bf Fig.1} Ratio of phase space integrals with and
   without momentum cutoff, for a $d$-wave neutron wave function.
   The effect of the cutoff is to include less than 90\%, between
   90\% and 95\% and more than 95\%  of the initial momentum distribution
   as marked on the figure.
   \vskip .2in
   \noindent {\bf Fig.2} Top left: Experimental cross sections for neutron scattering on
   $^{9}Be$ together with the optical model cross sections using the potential of Table I.
   The full curve is the total cross section, the dotdashed curve is the elastic cross
   section while the dashed
    curve is the reaction cross section. Bottom left: same quantities
   calculated in the eikonal approximation. Top and bottom right: the same
   quantities calculated with the Varner potential.

   \vskip .2in
   \noindent {\bf Fig.3} The behavior of the term $|1-S|^2$ and of the
   transmission coefficient $1-|S|^2$ calculated by the optical model
   (solid and dashed line respectively) and by the eikonal approximation
   (dotted and dotdashed line). Results are shown as a function of the
   angular momentum $l_n$. In the case of the eikonal calculations the
   semiclassical relation $l_n+1/2=kb_n$ was used to make the connection with the neutron
   impact parameter. Results are given at incident energies ranging from 20 to 100
   MeV. 

   \vskip .2in
   \noindent {\bf Fig.4}
   Total breakup probabilities from the 1/2$^-$ state in $^{12}Be$, as a function of the
   ion-ion impact parameter for diffraction obtained in the TC calculation by the solid
   line. The dashed line is for the stripping. Energies are
   as in Fig. 3. Dotted and dotdashed lines are diffraction and stripping probabilities
   from the eikonal calculation.

   \vskip .2in
   \noindent {\bf Fig.5}
   The integrand function of the diffraction (a), and stripping (b) term of Eq.(\ref{pg})
after $k_z$ integration, full curve, obtained
   from the realistic bound state wave function and the corresponding terms, diamonds,
   in the sum over partial waves of Eqs.(\ref{5},\ref{6}) in the
   case of the incident energy of 78A MeV. Crosses are the results of a calculation of Eq.
   (\ref{pg}) in which the eikonal phase shifts have been substituted by the
   optical model phase shifts according to Eq.(\ref{a10}). All calculations done at fixed
impact
   parameter
   $b_c=5.6fm$ between the projectile and target. 

\begin{figure}[htb]

\includegraphics{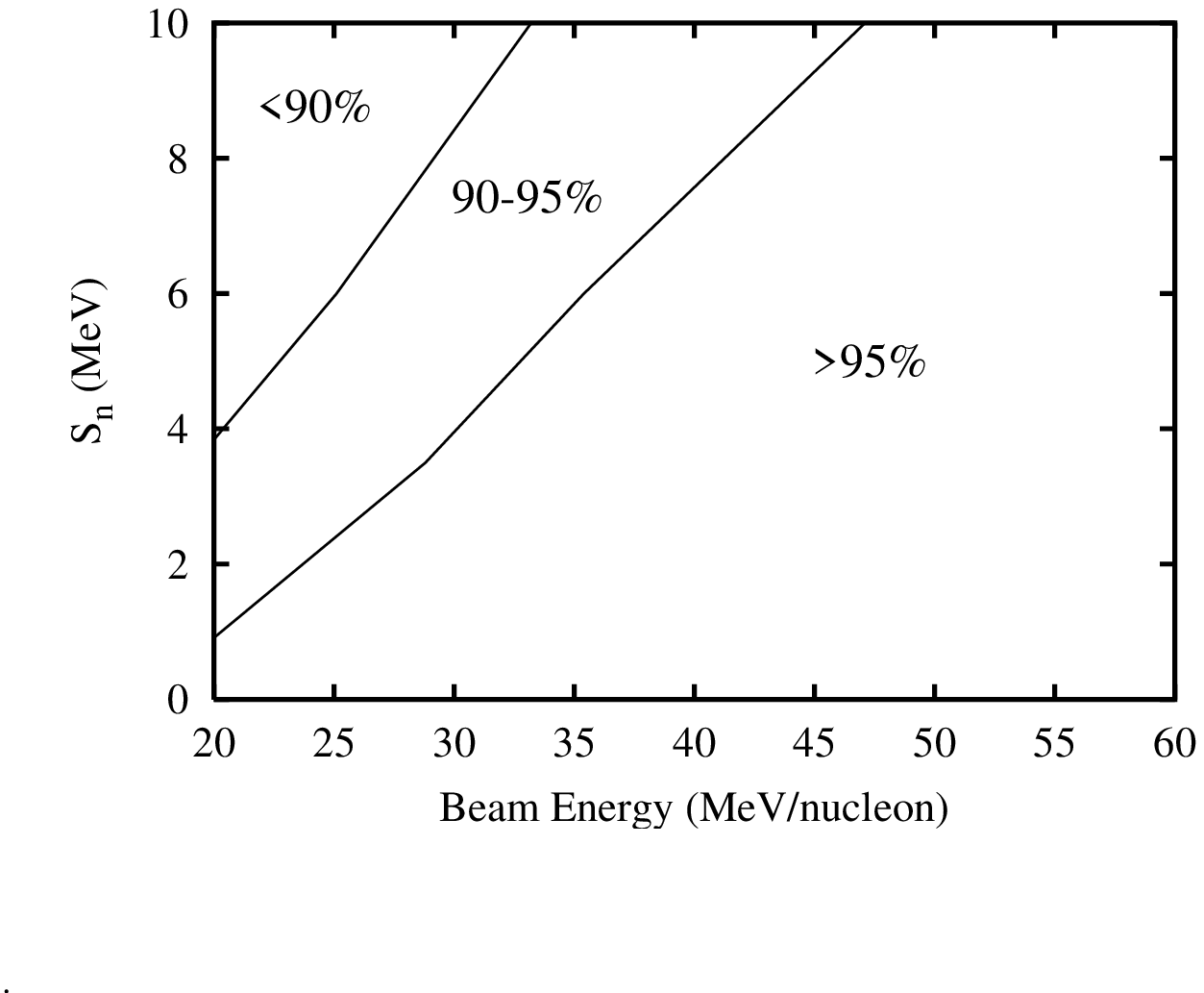}
\end{figure}
\begin{figure}[htb]

\includegraphics{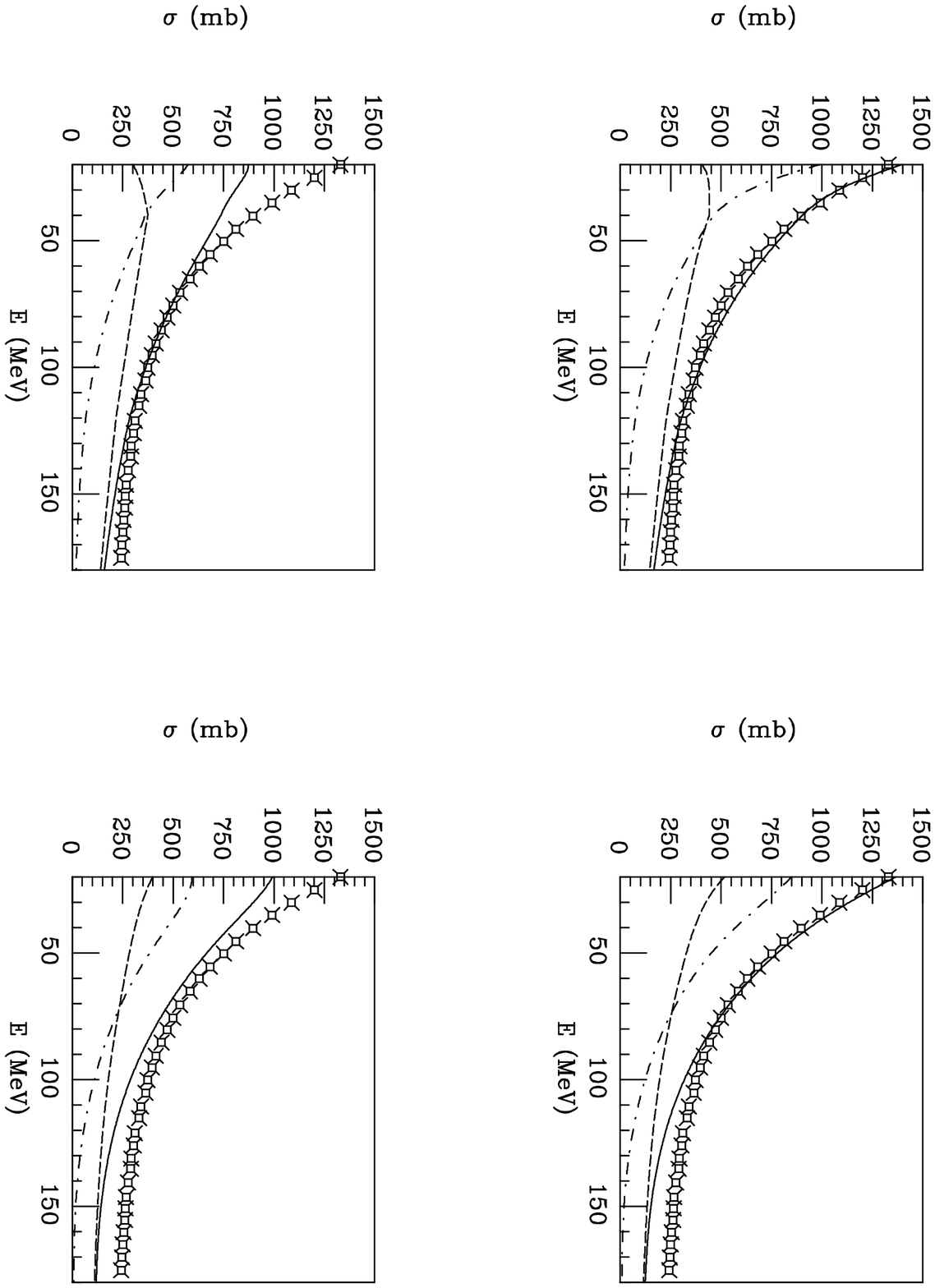}
\end{figure}
\begin{figure}[htb]

\includegraphics{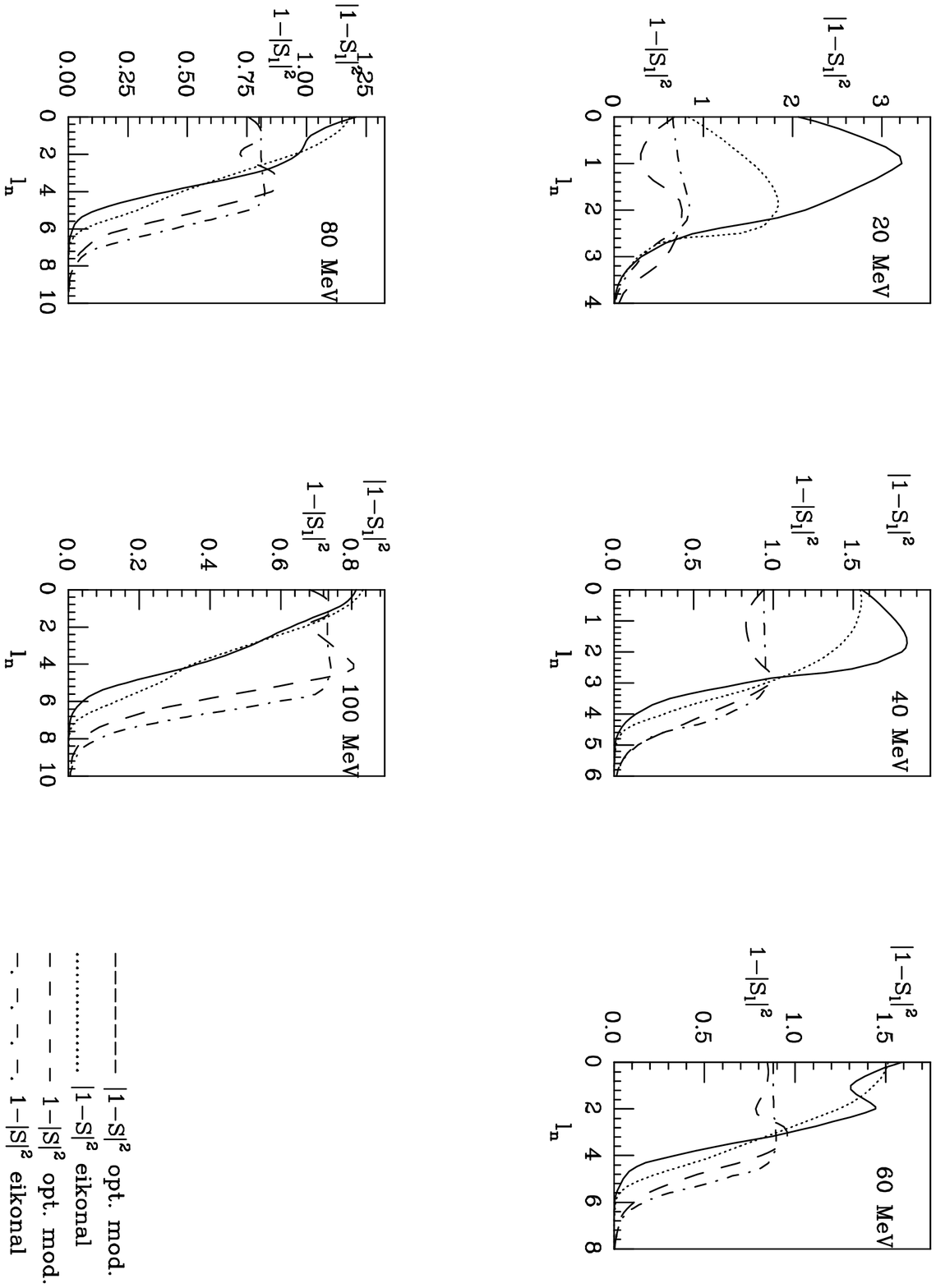}
\end{figure}
\begin{figure}[htb]

\includegraphics{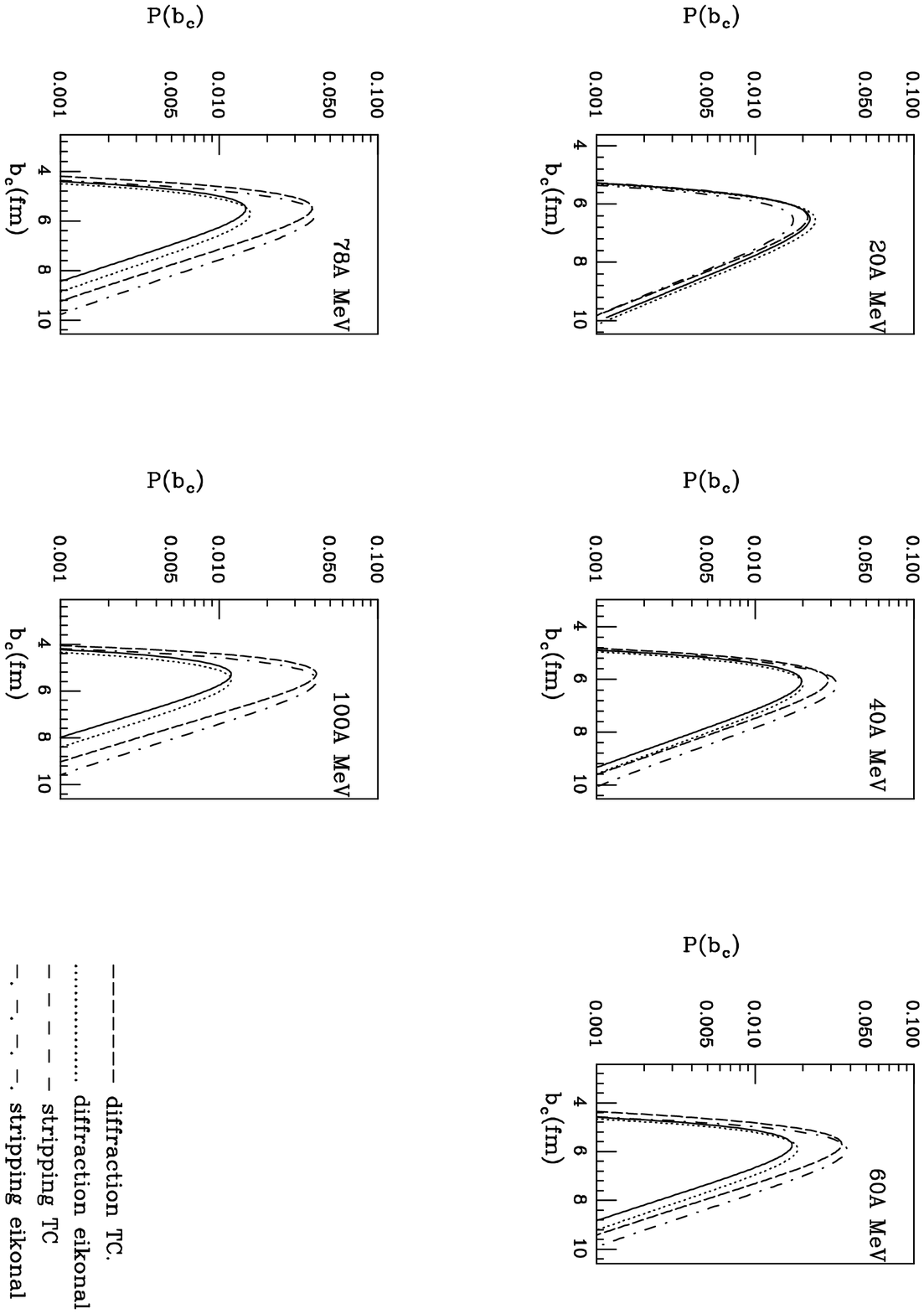}
\end{figure}
\begin{figure}[htb]

\includegraphics{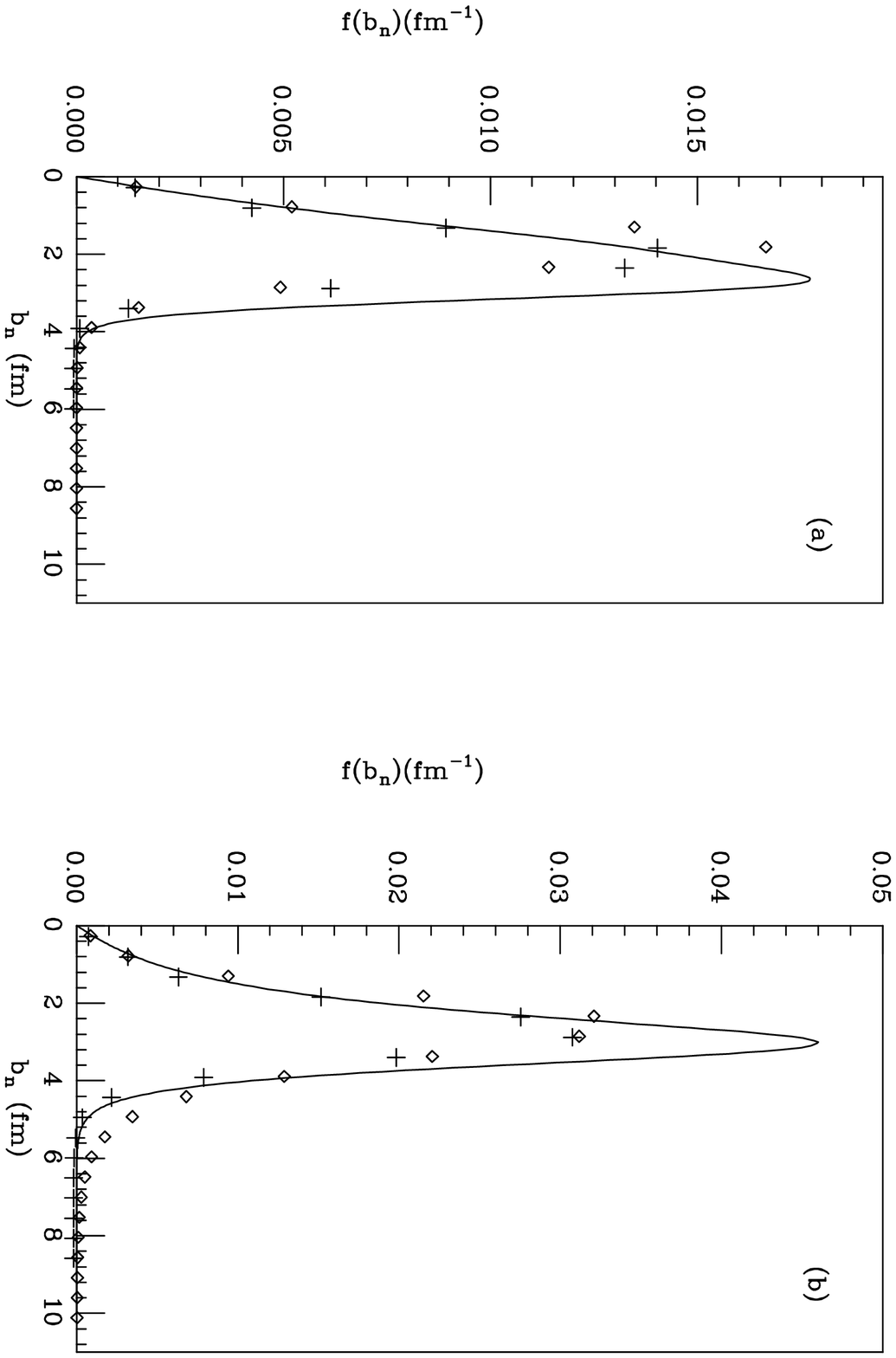}

\end{figure}

\newpage
   \begin{thebibliography}{40}


   \bibitem{yab} K. Yabana, Y. Ogawa and Y. Suzuki, Nucl. Phys. {\bf A539}, 295 (1992).

   \bibitem{esb} K. Hencken, G. F. Bertsch and H. Esbensen, Phys. Rev. C {\bf 54}, 3043
   (1996).

   \bibitem{bhe} G.F. Bertsch, K. Hencken and H. Esbensen, Phys. Rev. C {\bf 57}, 1366
   (1998).

 \bibitem{ab} A. Bonaccorso and D. M. Brink, Phys. Rev. C {\bf 57}, R22 (1998).

   \bibitem{jef} J. A. Tostevin, J. Phys. G {\bf 25}, 735 (1999).


   \bibitem{esb1} H. Esbensen and G. F. Bertsch, Phys. Rev. C {\bf 59}, 3240 (1999).

\bibitem{flo} F. Negoita et al., Phys. Rev. C {\bf 59}, 2082 (1999).

\bibitem{bb} A. Bonaccorso and D. M. Brink, Phys. Rev. C {\bf 38}, 1776 (1988).


   \bibitem{bb1} A. Bonaccorso and D. M. Brink, Phys. Rev. C {\bf 43}, 299 (1991).

   
   \bibitem{abb} A. Bonaccorso and D. M. Brink, Phys. Rev. C {\bf 58}, 2864 
   (1998).

  

   
   \bibitem{me} A. Bonaccorso, Phys. Rev. C {\bf 60}, 054604, (1999). 

   \bibitem{abfc} A. Bonaccorso and F. Carstoiou, Phys. Rev. C {\bf 61}
   034605, (2000).

   \bibitem{be61} M. A. B. ~B\'eg, Ann. Phys. 13, 110 (1961).

   \bibitem{sat} G. R. Satchler, {\it Direct Nuclear Reactions}, Clarendon
   Press, Oxford 1983. Pag 715.

   \bibitem{ker} J. Rapaport and A. K. Kerman, Nucl. Phys. {\bf A119}
   , 641 (1968).

   \bibitem{huf} T. Fujita and J. H\"ufner, Nucl. Phys. {\bf A343}, 493
   (1980).

   \bibitem{be12} A. Navin et al., Phys. Rev. Lett. {\bf 85}, 266 (2000).

   \bibitem{jlm} J. P. Jeukenne, A. Lejeune and C. Mahaux, 
   Phys. Rev. C {\bf 16}, 80 (1977).

   \bibitem{var} R. L. Varner, W. J. Thompson, T. L. McAbee, E. J. Ludwing and 
   T. B. Clegg, Phys. Rep. {\bf 201}, 57 (1991).



   \bibitem{pot1} J. H. Dave and C. R. Gould, Phys. Rev. C {\bf 28}, 2212 (1983).



   \bibitem{wat} B. A. Watson, P. P. Singh and R. E. Segel, Phys. Rev.
   {\bf 182}, 977 (1969).

   \bibitem{ma1} C. Mahaux, H. Ng\^ o and G. R. Satchler, Nucl. Phys. {\bf A449}, 354
   (1986).

   \bibitem{fin} R. W. Finlay, W. P. Abfalterer, G. Fink, E. Montei, T. Adami,
    P. W. Losowski, G. L.Morgan and R. C. Haight, Phys. Rev. C {\bf 47}, 237 (1993).

   \bibitem{ph} P. E. Hodgson, Phys. Lett. {\bf 65B}, 331 (1976).

   \bibitem{love} G. R. Satchler and W. G. Love, Phys. Rep. {\bf 55}, 
   183, (1979).

   \bibitem{av} C. E. Aguiar, F. Zardi and A. Vitturi, Phys. Rev. C {\bf
   56}, 1151 (1997).

  \bibitem{floc} F. Carstoiu, private communication.

\bibitem{jat} R. C. Johnson, J. S. Al-Khalili and J. A. Tostevin,
Phys. Rev. Lett. {\bf 79},  2771 (1997).

   \bibitem{tiina} A. Bonaccorso, I. Lhenry and T. S\"uomijarvi, Phys. Rev. C
   {\bf 49}, 329 (1994). 

   \bibitem{bb4} A. Bonaccorso , Phys. Rev. C {\bf 53}, 849 (1996).

   \bibitem{henri} H. Laurent et al. , Phys. Rev. C {\bf 52}, 3066 (1995).

   Elke Plankl, These, Universite Paris VII, June 1999, unpublished, report 
   IPNO-T-99-03.

    
   \end{thebibliography}
\end{document}